\input phyzzx.tex
\tolerance=1000
\voffset=-0.0cm
\hoffset=0.7cm
\sequentialequations
\def\rl{\rightline}

\def\t1{{\tilde 1}}

\REF{\DIM}{N. Arkani-Hamed, S. Dimopoulos and G. Dvali,  Phys. Rev. Lett.
{\bf B429} (1998) 263, hep-ph/9803315;
I.~Antoniadis, N.~Arkani-Hamed, S.~Dimopoulos and G.~Dvali,
Phys. Lett. {\bf B436}, 257 (1998), hep-ph/9804398.}
\REF{\OLD}{I. Antoniadis, Phys. Lett. {\bf B246} (1990) 377; J. Lykken,
 Phys. Rev. {\bf D54} (1996) 3693, hep-th/9603133; I. Antoniadis and K.
Benakli, Phys. Lett. {\bf B326} (1994) 69;
I. Antoniadis, K. Benakli and M. Quiros, Nucl. Phys. {\bf B331} (1994) 313.}
\REF{\EXP}{N. Arkani-Hamed, S. Dimopoulos and G. Dvali,  hep-ph/9807344.}
\REF{\OTH}{K.R.~Dienes, E.~Dudas and T.~Gherghetta,
Phys. Lett. {\bf B436}, 55 (1998)
hep-ph/9803466; R.~Sundrum, hep-ph/9805471, hep-ph/9807348; G.~Shiu and
S.H.~Tye,
Phys. Rev. {\bf D58}, 106007 (1998) hep-th/9805157;
P.C.~Argyres, S.~Dimopoulos and J.~March-Russell, Phys. Lett. {\bf B441}, 96
(1998)
hep-th/9808138; N. Arkani-Hamed, S. Dimopoulos and J. March-Russell,
hep-th/9809124;
A.~Donini and S.~Rigolin, hep-ph/9901443; Z. Berezhiani and G. Dvali,
 hep-ph/9811378; Z.~Kakushadze, hep-th/9812163, hep-th/9902080; Z. Kakushadze
and S.-H.H. Tye, hep-th/9809147; N. Arkani-Hamed, S. Dimopoulos,
hep-ph/9811353;
N. Arkani-Hamed, S. Dimopoulos, G. Dvali and J. March-Russell, hep-ph/9811448;
A.~Pomarol and M.~Quiros, Phys. Lett. {\bf B438}, 255 (1998);
T.E.~Clark and S.T.~Love, hep-th/9901103; T. Banks, M. Dine and A. Nelson,
hep-th/9903019;
I. Antoniadis, K. Benakli and M. Quiros, hep-ph/9905311; G. Shiu, R. Schrock
and S.-H. Tye, hep-ph/9904262; E. Halyo, hep-ph/9904432.}
\REF{\ACC}{G.F. Giudice, R. Rattazzi and J.D. Wells, hep-ph/9811291;
 T. Han, J.D. Lykken and R. Zhang, hep-ph/9811350;  J.L. Hewett,
hep-ph/9811356;
E.A. Mirabelli, M. Perlstein and M.E. Peskin,  hep-ph/9811337;  M.L.~Graesser,
hep-ph/9902310;
S. Nussinov and R.E. Shrock, hep-ph/9811323;
T.G. Rizzo, hep-ph/9901209; hep-ph/9902273; hep-ph/9903475;   K.~Agashe and
N.G.~Deshpande, hep-ph/9902263; K. Cheung, hep-ph/9904266; K. Cheung and W.
Keung,
hep-ph/9903294;
 M.~Masip and A.~Pomarol, hep-ph/9902467.}
\REF{\COS}{K. Benakli, hep-ph/9809582; K. Benakli and S. Davidson,
hep-ph/9810280; M. Maggiore and A. Riotto, hep-th/9811089; D.H. Lyth,  Phys.
Lett. {\bf B448} (1999) 191, hep-ph/9810320; N. Kaloper and A. Linde,
hep-th/9811141; G. Dvali and S.-H. H. Tye, hep-ph/9812483; K.R. Dienes, E.
Dudas, T. Ghergetta, A. Riotto,
hep-ph/9809406; N. Arkani-Hamed, S. Dimopoulos, N. Kaloper and J.
March-Russell, hep-ph/9903224; hep-ph/9903239; L. Hall and D. Smith,
hep-ph/990426; E. Halyo, hep-ph/9901302; hep-ph/9905244;
C. Csaki, M. Graesser and J. Terning, hep-ph/9903319; J. Cline, hep-ph/9904495;
A. Riotto, hep-ph/9904485; G. Dvali, hep-ph/9905204.}
\REF{\CUL}{S. Cullen and M. Perelstein, hep-ph/9903422.}
\REF{\UNI}{K.R.~Dienes, E.~Dudas and T.~Gherghetta, hep-ph/9806292; C. Bachas,
hep-ph/9807415;  Z. Kakushadze, hep-ph/9811193; I. Antoniadis and C. Bachas,
hep-th/9812093; A. Perez-Lorenzana and R. Mohapatra, hep-ph/9904504.}
\REF{\IBA}{L. Ibanez, hep-ph/9505349.}
\REF{\SIG}{L. Ibanez, R. Rabadan and A. Uranga, hep-th/9905098.}
\REF{\AFIV}{G. Aldazabal, A. Font, L. Ibanez and G. Violero, hep-th/9804026.}
\REF{\IMR}{L. Ibanez,  C. Munoz and S. Rigolin, hep-ph/9812397.}
\REF{\IRU}{L. Ibanez, R. Rabadan and A. Uranga, hep-th/9808139.}

\singlespace
\rl{SU-ITP-99-26}
\rl{hep-ph/9905577}
\rl{\today}
%\rl{T}
\pagenumber=0
\normalspace
\medskip
\bigskip
\titlestyle{\bf{ Mirage Gauge Coupling Unification and TeV Scale Strings}}
\smallskip
\author{ Edi Halyo{\footnote*{e--mail address: halyo@dormouse.stanford.edu}}}
\smallskip
\centerline {Department of Physics}
\centerline{Stanford University}
\centerline {Stanford, CA 94305}
\smallskip
\vskip 2 cm
\titlestyle{\bf ABSTRACT}

We consider gauge coupling unification in models with TeV scale strings and
large compact dimensions realized as type IIB string orientifolds.
Following an observation by Ibanez we show that the gauge couplings at low
energies can behave as if they effectively unify at $M_U \sim 2 \times
10^{16}~GeV$ with $\alpha_U \sim 1/24$. This requires the $\sigma$ model
anomaly coefficients $b_a^{i \prime}$  not to be all equal and their ratio to
the $\beta$--functions of minimally supersymmetric Standard Model $\beta_a$ to
be a constant independent of the gauge group. If, in addition, $b_a^{i \prime}$
have a gauge group independent constant piece the relation between the unified
gauge coupling and the dilaton VEV is modified so that there can be weakly
coupled gauge theories arising from strongly coupled strings.

\singlespace
\vskip 0.5cm
\endpage
\normalspace

\centerline{\bf 1. Introduction}
\medskip

Lately there has been great interest in scenarios involving TeV scale strings
compactified on two or more large dimensions[\DIM,\OLD]. In these models the
Standard Model degrees of freedom are confined to the world--volume of
D--branes. Gravity propagates only in the bulk and its weakness compared to
other interactions is a result of the large compact dimensions
$$M_P^2={8 M_s^8 R^6\over {g_s^2}} \eqno(1)$$
Here $g_s$ and $M_s$ are the string coupling and string scale respectively and
$R^6$ is the generic volume of the compact dimensions. Surprisingly, this
scenario cannot be easily ruled out by
precision and acceleretor experiments or astrophysical observations[\EXP].
Implications of this scenario for particle phenomenology, astrophysics and
cosmology  have been investigated
in some detail[\OTH,\ACC,\COS]. The strongest bound
on  the higher dimensional Planck  scale, $M_{d+4}$ comes from the supernova
1987A data and is $M_6>50~TeV$ for two large dimensions and $M_{d+4}>1~TeV$ for
more than two large dimensions[\CUL]. In this letter, for simplicity, we take
the bound on the higher dimensional Planck scale to hold over the string scale
$M_s$.

On the other hand, one of the most attractive results of minimally
supersymmetric Standard Model  with the assumption of a desert up to very high
scales is the unification of
the gauge couplings with $\alpha_U=g_U^2/4\pi \sim 1/24$ at $M_U \sim 2 \times
10^{16}~GeV$.
In models with TeV scale strings, around $M_s$ we find
a plethora of new states such as excited string modes, Kaluza--Klein modes,
winding modes etc.
Moreover field theory is not applicable at this scale and one should use
fully--fledged string theory.
Therefore unification of gauge couplings at very high energy scales such as
$M_U$ would seem to be completely accidental. Some aspects of gauge coupling
unification in the framework of TeV scale strings and large compact dimensions
have been investigated in [\UNI].

Recently Ibanez has pointed out that in $D=4$, $N=1$ supersymmetric IIB string
orientifold models there is a possibility of obtaining such effective or mirage
gauge coupling unification at scales which are much higher than the string
scale[\IBA,\SIG]. These models generically have $\sigma$ model
anomalies in addition to anomalous $U(1)$ gauge groups. All anomalies are
cancelled by the
Green--Schwarz mechanism which results in moduli dependent corrections to the
gauge couplings. When ratio between the coefficients of these corrections
($\sigma$ model anomaly coefficients) and the $\beta$--functions are constants
independent of the gauge group
these corrections can be absorbed into the effective unification scale. For
models with large compact dimensions the effective unification scale becomes
much larger than the string scale. Since there is no running of gauge couplings
above $M_s$ and there is a fully--fledged string theory at this scale,
this unification is a mirage seen from the low--energy point of view. In fact
the gauge couplings do not unify at the string scale.

In this letter we show that in TeV scale string scenarios realized by $D=4$,
$N=1$ supersymmetric IIB string orientifold
models an effective unification of couplings at $M_U$ is possible. For this to
occur it is important to have
an anisotropic compactification such as two large and four string size compact
dimensions.
Moreover, as we explain below, the $\sigma$ model anomaly coefficients must not
all be equal to each other, preferably one much larger than the others. If in
addition, these coefficients contain a constant term independent of the gauge
group then the well--known relation between the unified gauge coupling and the
dilaton VEV is modified. In this case a string model with intermediate or
strong coupling can lead to weakly coupled gauge theories.

\bigskip
\centerline{\bf 2. Mirage Gauge Coupling Unification}
\medskip

In this section we consider a simplified model which has all the
characteristics of  $D=4$, $N=1$ supersymmetric IIB string orientifold models
with large internal dimensions[\AFIV,\IMR,\IRU]. These models are obtained by
compactifying the IIB string on $T^6=T^2 \times T^2 \times T^2$ and moding out
by
world--sheet parity $\Omega$ times a discrete space--time symmetry. The models
generically have a chiral spectrum with a large gauge group. Part of the
massless spectrum is projected out and the gauge group is broken by Wilson
lines. We will assume that a realistic string model can be constructed in this
framework even though
such a model does not exist yet. Depending on the orientifold group
there will be 32 D--branes of a given dimension (such as D3 or D5 branes) with
64 orientifold planes with the same dimension. Gauge bosons and matter live on
the brane world--volumes whereas gravity only propagates in the bulk. For more
details on these models we refer the reader to the refs. [\AFIV,\IMR,\IRU].
Below
we consider a toy model with all the relevant properties of these string
models.

For simplicity, we assume that the gauge group is the Standard Model group
$SU(3)_C \times SU(2)_L \times
U(1)_Y $ and another $U(1)_X$ which is anomalous. This anomaly is cancelled by
the Green--Schwarz mechanism. The Kahler potential for the dilaton $S$, the
untwisted moduli  $T_i$ and the matter fields $\phi_r$ are given by
$$K(S,S^*,T_i,T_i^*,\phi_r,\phi_r^*)=-log(S+S^*)-\sum_i
log(T_i+T_i^*)+\sum_{r,i}{\phi_r \phi_r^* \over{(T_i+T_i^*)^{n_r^i}}}
\eqno(2)$$
where $n_r^i$ are the modular weights of the matter fields $\phi_r$ with
$\Sigma_i n_r^i=1$.
The gauge function for the group denoted by index $a$ ($a=1,2,3$) is given by
$$f_a=S+{\delta_a \over 2}M \eqno(3)$$
where $M$ is the overall twisted modulus of the model and $\delta_a$ is a
constant
which depends only on the gauge group.
This model has four anomalies; the $U(1)_X$ anomaly and three $SL(2,R)_i$
$\sigma$ model anomalies due to the three $SL(2,R)_i$ transformations[\SIG]
$$T_i \to {{a_iT_i-ib_i} \over {ic_iT_i+d_i}} \eqno(4)$$
with $a_i,b_i,c_i,d_i$ real and $a_id_i-b_ic_i=1$.
All the anomalies are cancelled by the Green--Schwarz mechanism if the
imaginary part of the twisted modulus transforms as ($\Lambda_X$ is the
$U(1)_X$ gauge parameter)
$$Im M \to Im M+\delta^X_{GS} \Lambda_X \eqno(5)$$
under the $U(1)_X$ and as
$$Im M \to Im M-2\sum_i \delta_{GS}^i log (icT_i+d) \eqno(6)$$
under the three $SL(2,R)_i$ model transformations.
Here $\delta_{GS}^X$ and $\delta_{GS}^i$ are the anomaly coefficients of the
$U(1)_X$ and
$\sigma$ model anomalies respectively.
We also assume that the twisted modulus has the Kahler potential
$$K(M,M^*)=(M+M^*-\delta^X_{GS} V_X-\sum_i  \delta_{GS}^i log(T_i+T_i^*))^2
\eqno(7)$$
which is invariant under both $U(1)_X$ and the $\sigma$ model transformations.
The above Kahler potential induces an anomalous D--term[\IBA,\SIG]
$$\xi_X=-\delta^X_{GS} (M+M^*-2\sum_i  \delta_{GS}^i log(T_i+T_i^*)) \eqno(8)$$
If we are interested in supersymmetric vacua then $\xi_X=0$ which means
$$ReM = \sum_i  \delta_{GS}^i log(T_i+T_i^*) \eqno(9)$$
As a result, the gauge couplings are given by
$${8\pi^2 \over g_a^2}= ReS+{1 \over 2} \sum_i (\delta_{GS}^i \delta_a)
log(T_i+T_i^*) \eqno(10)$$
The product of the two anomaly coefficients is $\delta_{GS}^i \delta_a=b_a^{i
\prime}$ where[\SIG]
$$b_a^{i \prime}=-C(G_a)+\sum_{R_a} T(R_a)(1+2n^i_{R_a}) \eqno(11)$$
Here $C(G_a)$ and $T(R_a)$ are the second Casimir invariants of the adjoint and
$R_a$ representations of the gauge group $G_a$. Moreover, it can be shown that
$\Sigma_i b_a^{i \prime} =\beta_a$ where $\beta_a$ are the $\beta$--functions
corresponding to the minimally supersymmetric standard model.

The running gauge couplings become[\IBA]
$${8\pi^2 \over g_a^2}= ReS+{1 \over 2} \sum_i  b_a^{i \prime} log(T_i+T_i^*)
+{1 \over 2} \beta_a log{M_s^2 \over Q^2} \eqno(12)$$
Here we assume that there are no threshold states including Kaluza--Klein
states up to the string scale. This is only possible if the Standard Model
degrees of freedom arise from the world--volume of D3 branes so that there are
no brane directions wrapped around string size compact dimensions.
We see that the correction due to the $\sigma$ model anomaly
gives an extra moduli dependent logarithmic term in the running coupling
constant.
If  $b_a^{i \prime}= r_i \beta_a$ with $r_i$ a constant independent of the
gauge group then
we see that the moduli dependent corrections can be combined with $M_s$ to give
an
effective unification scale
$$M_X=M_s \prod _i (T_i+T_i^*)^{r_i/2} \eqno(13)$$
The untwisted moduli $T_i$ are related to the compactification radii by
$$2ReT_i={4 R_j^2 R_k^2 M_s^4 \over g_s} \eqno(14)$$
Therefore we get
$$M_X={2M_s^3 \over {g^{3/2}}} ( R_j R_k )^{r_i} ( R_i R_k)^{r_j} ( R_i
R_j)^{r_k} \eqno(15)$$

We have found that due to the $\sigma$ model anomaly there is an extra moduli
dependent logarithmic term in the running coupling constant. Under certain
conditions i.e. $b_a^{i \prime} =r_i
\beta_a$ for all gauge groups, this term
modifies the unification scale. We see that effectively the unification scale
becomes $M_X$ with the unified gauge coupling $g_U$ given by $ReS$. In models
with large compct dimensions the scale $M_X$ is much larger than the string
scale $M_s$. However, we know that the gauge couplings do not run above the
string scale, $M_s$, in fact above $M_s$ we have a fully--fledged string
theory.
Therefore, the field theoretical running of the gauge couplings up to $M_X$ and
their unification at $M_X$ are just  mirages seen from the low energies.
Note that at the string scale the gauge couplings are not unified but given by
eq. (12) with $Q^2=M_s^2$.

In fact we can go further and assume that $b_a^{i \prime}= r_i \beta_a+\eta_i$
where $\eta_i$ is a constant independent of the gauge group. Then we find that
in addition to the modification of the unification scale, there is a constant
term which can be absorbed into the dilaton VEV.
We have
$$ {8\pi^2 \over g_U^2}= ReS+\sum_i \eta_i log (T_i+T_i^*) \eqno(16)$$
This result is very interesting. If $\eta_i=0$ then the unified gauge coupling
$g_U$ is fixed by $ReS$. In particular since $\alpha_U=1/24$ we need $ReS \sim
150$ a large value
exactly as in heterotic string theory. Since $ReS=2/g_s$ this also corresponds
to a weakly coupled string theory.
However, if $\eta_i \not= 0$, $ReS$ does not need to correspond to the unified
gauge coupling. In particular, it can be much smaller than $150$ or even of
$O(1)$ depending on $\eta_i$ and the size of the compact dimensions. In this
case there is the possibility of having weakly coupled gauge interactions in a
string theory with intermediate (or strong) i.e. $O(1)$ coupling.

Our aim is to find out whether an effective mirage unification of the gauge
couplings can be
obtained with $\alpha_U \sim 1/24$ and $M_X \sim 2 \times 10^{16}~GeV$ in this
framework.
We first assume that $\eta_i=0$ and the compactification is isotropic so that
there is only one overall untwisted modulus $T=T_1+T_2+T_3$. Then using
$\Sigma_i b_a^{i \prime} =\beta_a$
we find
$$M_X= {2 R^2 M_s^3 \over {g_s^{3/2}}} \eqno(17)$$
where the compactification radius $R$ is given by
$$R^6={g_s^2M_P^2 \over {8M_s^8}} \eqno(18)$$
As observed in ref. [\IBA] this gives $M_X \sim 10^{13}~GeV$ which is three
orders of magnitude smaller than
the desired value. Of course, this example is too simple because in models with
large internal dimensions, the compactification is not isotropic and one should
consider the three untwisted moduli $T_i$ separately.

We now consider TeV scale strings compactified on two large and four string
size dimensions.
The cases with more than two large dimensions can be easily generalized from
this case and
work as well.
The size of the large dimensions is given by
$$R^2={g_U^4 M_P^2 \over {32\pi^2 M_s^4}} \eqno(19)$$
If the two large dimensions are on the same torus we have
$$2ReT_1=2ReT_2={g_U^2 \over {4 \pi}} \left({M_P \over M_s}\right)^2
\eqno(20)$$
and
$$2ReT_3={8\pi \over g_U^2} \eqno(21)$$
We now need to assume values for $b_a^{i'}$ which need to be proportional to
$\beta_a$ independently of $a$.
We take for example $r_i=(3/4,1/8,1/8)$ for the three tori $T^2_i$.  We stress
that this choice is the same for the three gauge groups $SU(3)_C,
SU(2)_L,U(1)_Y$. Then
$$M_X \sim \alpha_U^{5/16} M_P^{7/8} M_s^{1/8} \eqno(22)$$
For $M_P \sim 10^{19}~GeV$, $M_s \sim 50~TeV$ and $g_U^2 \sim 1/2$ we find
$M_X \sim 2 \times 10^{16}~GeV$ which is the desired unification scale
predicted by the
minimally supersymmetric Standard Model. We see that the dependence on the
string scale $M_s$ is very weak so that a change of $M_s$ between $1~TeV$ and
$50~TeV$ gives only a factor of $(50)^{1/8} \sim 1.5$. This means that mirage
unification is not too sensitive to
the lowest bound on the string scale, $M_s$ or to the distinction between $M_s$
and the higher dimensional Planck scale $M_{d+4}$.
Also note that for the above choice of $b_a^{i \prime}$ either the matter
content of the Standard Model arises in an asymmetric manner from the sectors
corresponding to the three tori or the modular weights $n_r^i$ are not the same
in these sectors.

The choice above for $b_a^{i \prime}$ is a very unique one and it is not clear
why the anomaly coefficients should satisfy this. However, we remind that the
above holds only for $\eta_i=0$. If $\eta_i \not=0$, we can change each one of
the anomaly coefficients by a constant which does not depend on the gauge
group. Thus, there are many values
of $b_a^{i'}$ that will result in mirage gauge coupling unification. The only
requirement on the coefficients $b_a^{i'}$ seems to be that they be the same
for all gauge groups for a given torus.

On the other hand, when $\eta_i \not=0$ the value of the unified gauge coupling
does not arise solely from
$ReS$ as given in eq. (10). In order to estimate the magnitude of this effect
we assume that
$\eta_1 \not=0$ but $\eta_2=\eta_3=0$. Then
$$ {8\pi^2 \over g_U^2}= ReS+ \eta_1 log (T_1+T_1^*) \eqno(23)$$
Using eq. (20) we find that the correction to $ReS$ is $\sim 30 \eta_1$. So
even for $\eta_1 \sim 1$
this is an important effect. Of course this effect is even larger when all
$\eta_i \not =0$.
If $\eta_1 \sim 5$, $ReS$ can be of $O(1)$ without changing the unified gauge
coupling $g_U$.
Thus, we find the interesting possibility of getting weakly coupled gauge
theories from a string theory which has an intermediate or strong coupling.

\bigskip
\centerline{\bf 3. Conclusions and Discussion}
\medskip

In this letter we showed that an effective unification of gauge couplings at
$M_U \sim 2 \times 10^{16}~GeV$ can be obtained in models with TeV scale
strings. In models with $\sigma$ model anomalies the expression for the gauge
couplings get moduli dependent corrections. If the anomaly coefficients are
proportional to the $\beta$--functions independently of the gauge group
there is an effective new unification scale. For models with large compact
dimensions this effective unification scale is much larger than $M_s$. However,
at $M_s$ there is fully--fledged string theory and therefore gauge couplings do
not run above this scale.
The field theoretical running
of the gauge couplings up to energy scales much larger than $M_s$ and their
unification at
$M_U$ is a mirage from the low energy point of view.
For simplicity we considered only the case with two large compact dimensions
but our results can be easily generalized to the cases with more than two large
dimensions.

In order for mirage unification to occur for two large dimensions  the anomaly
coefficients have to satisfy $b_a^{i \prime}=r_i \beta_a$ with
$r_i=(3/4,1/8,1/8)$ for all the gauge groups. Therefore either the observable
sector must arise in an asymmetric manner from the sectors corresponding to the
three tori or the modular weights must be asymmetric among these sectors. This
is a unique choice for $b_a^{i \prime}$ which is not generic at all. However,
unification can be maintained if the anomaly coefficients satisfy the modified
relation
$b_a^{i \prime}=r_i \beta_a+\eta_i$. Thus, there are a large number of choices
for $b_a^{i \prime}$ which result in mirage unification. It is difficult to say
how generic this situation is because there are no realistic $D=4$, $N=1$
supersymmetric IIB orientifold models. The requirement  $r_i$ have
to satisfy for mirage unification is that they be the same for all gauge groups
and not be equal to each other. It seems that demanding mirage gauge coupling
unification puts strong constraints on TeV scale string model building.
At this stage, it is hard to say whether these can be easily
satisfied.

An intriguing result of the expression $b_a^{i \prime}=r_i \beta_a+\eta_i$ is
the change in the relation between the value of the unified coupling constant
and the dilaton VEV. From eq. (10) we see that for some values of  $\eta_i$ and
large compact dimensions, the gauge couplings can be small even though the
string coupling $g_s=2/ReS$ is large. Therefore a string theory with an
intermediate or
strong coupling can give rise to perturbative gauge interactions.
This may facilitate dilaton stabilization in TeV scale string models since now
there is no constraint on the dilaton VEV.

\refout
\vfill

\end
\bye